\documentclass[copyright,creativecommons,noderivs]{eptcs}
\def\event{The 21st Workshop on Logic-based methods in Programming\\ Environments}

\usepackage{breakurl}
\usepackage{pstricks}
\usepackage{underscore}
\usepackage{xspace}

\title{Profiling parallel Mercury programs \\ with ThreadScope}
\def\titlerunning{Profiling parallel Mercury programs with ThreadScope}
\author{Paul Bone%
    \thanks{Paul's work was supported by an Australian Postgraduate Award
       and a NICTA top-up scholarship.}~
  and Zoltan Somogyi
  \institute{Department of Computer Science and Software Engineering \\
             University of Melbourne, and}
  \institute{National ICT Australia (NICTA)}
  \email{\{pbone,zs\}@csse.unimelb.edu.au}
}
\def\authorrunning{Paul Bone and Zoltan Somogyi}

\newcommand{\authornote}[3]{
}

\newcommand{\tr}[1]{}

\begin{document}

\maketitle

\begin{abstract}
The behavior of parallel programs is even harder to understand
than the behavior of sequential programs.
Parallel programs may suffer from
any of the performance problems afflicting sequential programs,
as well as from several problems unique to parallel systems.
Many of these problems are quite hard
(or even practically impossible)
to diagnose without help from specialized tools.
We present a proposal for a tool
for profiling the parallel execution of Mercury programs,
a proposal whose implementation we have already started.
This tool is an adaptation and extension of the ThreadScope profiler
that was first built to help programmers visualize
the execution of parallel Haskell programs.
\end{abstract}

\section{Introduction}

When programmers need to improve the performance of their program,
they must first understand it.
The standard way to do this is to use a profiler.
Profilers record performance data from executions of the program,
and give this data to programmers
to help them understand how their program behaves.
This enhanced understanding then makes it easier
for programmers to speed up the program.

Profilers are needed because the actual behavior of a program
often differs from the behavior imagined by the programmer.
This is true even for sequential programs.
For parallel programs,
the gulf between human estimates and machine realities
is usually even wider.
This is because every cause
that can cause such differences for sequential programs
is present in parallel programs as well,
while parallel programs also have several causes unique to them.

Consider a program containing a loop with many iterations
in which the iterations do not depend on one another
and can thus be done in parallel.
Actually executing every iteration in parallel
may generate an overwhelming number of parallel tasks.
It may be that the average amount of computation done by one of these tasks
is less than the amount of work it takes to spawn one of them off.
In that case, executing all the iterations in parallel
will increase overheads to the point of actually slowing the program down,
perhaps quite substantially.
The usual way to correct this is to use granularity control
to make each spawned-off task execute
not one but several iterations of the loop,
thus making fewer but larger parallel tasks.
However, if this is done too aggressively,
performance may once again suffer.
For example,
there may be fewer tasks created than the computer has processors,
causing some processors to be idle when they could be used.
More commonly, the iterations of the loop and thus the tasks
may each need a different amount of CPU time.
If there are eight processors and eight unevenly sized tasks,
then some processors will finish their work early,
and be idle while waiting for the others.

All these problems can arise in programs with independent parallelism:
programs in which parallel tasks do not need to communicate.
Communication between tasks makes this situation even more complicated,
especially when a task may block waiting for information from another task.
A task that blocks may in turn further delay
other tasks that depend on data \emph{it} produces.
Chains of tasks that produce and consume values from one another are common,
Programmers need tools to help them
identify and understand performance problems,
whether they result from such dependencies or other causes.

Profiling tools that can help application programmers
can also be very useful for the implementors of runtime systems.
While optimizing Mercury's parallel runtime system,
we have needed to measure the costs of certain operations
and the frequency of certain behaviors.
Some examples are:
when a piece of work is made available,
how quickly can a sleeping worker-thread respond to this request?
When a task is made runnable after being blocked,
how often will it be executed on the same CPU that was previously executing it,
so that its cache has a chance to be warm?
Such information from profiles of typical parallel programs
can be used to improve the runtime system,
which can help improve the performance of \emph{all} parallel programs.

A third category of people who can use profiling tools for parallel programs
is researchers working on automatic parallelization tools.
Doing a good job of automatically parallelizing a program
requires a cost-benefit analysis for each parallelism opportunity,
which requires estimates of both the cost and the benefit of each opportunity.
Profiling tools can help researchers calibrate
the algorithms they use to generate estimates of both costs and benefits.

As researchers working on the parallel implementation of Mercury~\cite{jlp},
a pure logic programming language designed for
the creation of large, reliable and efficient application programs,
we fall into all three of the above categories.
We have long needed a tool
to help us understand the behavior of parallel Mercury programs,
of the parallel Mercury runtime system
and of our autoparallelization tool \cite{overlap},
but the cost of building such a tool seemed daunting.
We therefore looked around for alternative approaches.
The one we selected was to adapt to Mercury
an existing profiler for another parallel system.

The profiler we chose was ThreadScope~\cite{threadscope},
a profiler built by the team behind the Glasgow Haskell Compiler (GHC)
for their parallel implementation of Haskell.
We chose ThreadScope because
the runtime system it was designed to record information from
has substantial similarities to the Mercury runtime system,
because it was \emph{designed} to be extensible,
and, like many visualization tools, it is extremely useful
for finding subtle issues that cause significant problems.

The structure of the paper is as follows.
Section~\ref{sec:background} gives the background
needed for the rest of the paper:
it introduces Mercury,
the machinery involved in the parallel execution of Mercury programs,
and the ThreadScope profiler.
Section~\ref{sec:newevents}
describes how we extended the ThreadScope system
to collect the kinds of data needed to describe
the parallel execution of Mercury programs,
while section~\ref{sec:analysis} describes how
one can analyze that data to yield insights
that can be useful to our three audiences:
application programmers,
runtime system implementors,
and the implementors of autoparallelization tools.
Section~\ref{sec:conc}
concludes with a discussion of related work.

\section{Background}
\label{sec:background}

\subsection{Mercury}
\label{sec:backmer}

Mercury is a pure logic programming language.
The atomic constructs of Mercury are unifications,
plain first-order calls,
and higher-order calls.
Its composite constructs are
sequential and parallel conjunctions, disjunctions, switches,
if-then-elses, negations and existential quantifications.
A switch is a disjunction in which
each disjunct unifies the same bound variable
with a different function symbol;
the others should all be self-explanatory.

Mercury has a strong mode system.
The mode system classifies each argument of each predicate
as either input or output;
there are exceptions, but they are not relevant to this paper.
If input, the caller must pass a ground term as the argument.
If output, the caller must pass a distinct free variable,
which the predicate will instantiate to a ground term.
It is possible for a predicate to have more than one mode;
we call each mode of a predicate a \emph{procedure}.
The compiler generates separate code
for each procedure of a predicate.
The mode checking pass of the compiler is responsible for
reordering conjuncts (in both sequential and parallel conjunctions)
as necessary to ensure that for each variable shared between conjuncts,
the goal that generates the value of the variable (the \emph{producer})
comes before all goals that use this value (the \emph{consumers}).
This means that for each variable in each procedure,
the compiler knows exactly where that variable is bound to a ground term.

Each procedure and goal has a determinism,
which may put upper and lower bounds on the number of its possible solutions
(in the absence of infinite loops and exceptions).
A determinism may impose an upper bound of one solution,
and it may impose a lower bound of one solution.
\emph{det} procedures succeed exactly once;
\emph{semidet} procedures succeed at most once;
\emph{multi} procedures succeed at least once;
\emph{nondet} procedures may succeed any number of times.

\subsection{Parallelism in Mercury}
\label{sec:backpar}

The Mercury runtime system has a construct called a Mercury \emph{engine}
that represents a virtual CPU.
Each engine is independently schedulable by the OS, usually as a POSIX thread.
The number of engines that a parallel Mercury program will allocate on startup
is configurable by the user,
but it defaults to the actual number of CPUs.
Another construct in the Mercury runtime system is a \emph{context},
which represents a computation in progress.
An engine may be idle, or it may be executing a context;
a context can be running on an engine, or it may be suspended.
When a context finishes execution,
its storage is put back into a pool of free contexts for later reuse.
Following \cite{multicorerts},
we use \emph{sparks} to represent goals that have been spawned off
but whose execution has not yet been started.

The only parallel construct in Mercury is parallel conjunction:
$(G_1~\&~\ldots~\&~G_n)$.
All the conjuncts must be deterministic,
that is, they must all have exactly one solution.
This restriction greatly simplifies the implementation,
since it guarantees that there can never be any need
to execute $(G_2~\&~\ldots~\&~G_n)$ multiple times,
just because $G_1$ has succeeded multiple times.
(Any local backtracking inside $G_1$ will not be visible to the other conjuncts;
bindings made by det code are never retracted.)
In practice, we have not found this to be a significant limitation.

The Mercury compiler implements $(G_1~\&~G_2~\&~\ldots~\&~G_n)$
by creating a data structure representing a barrier,
and then spawning off $(G_2~\&~\ldots~\&~G_n)$ as a spark.
Since $(G_2~\&~\ldots~\&~G_n)$ is itself a parallel conjunction,
it is handled the same way:
the context executing it
first spawns off $(G_3~\&~\ldots~\&~G_n)$ as a spark,
and then executes $G_2$ itself.
Eventually, the spawned-off remainder of the conjunction
consists only of the final conjunct, $G_n$,
and the context just executes it.
The code of each conjunct synchronizes on the barrier once it has
completed its job.
When all conjuncts have done so,
the original context will continue execution after the parallel conjunction.

Mercury's mode system allows a parallel conjunct to consume variables
that are produced by conjuncts to its left, but not to its right.
This guarantees the absence of circular dependencies
and hence the absence of deadlocks between the conjuncts,
but it does allow a conjunct to depend on data that is yet to be computed
by a conjunct running in parallel.
We handle these dependencies through a source-to-source transform
\cite{wangdepparconj}.
The compiler knows which variables
are produced by one parallel conjunct and consumed by another.
For each of these shared variables,
it creates a data structure called a \emph{future} \cite{multilisp}.
When the producer has finished computing the value of the variable,
it puts the value in the future and signals its availability.
When a consumer needs the value of the variable,
it waits for this signal (if it has not yet happened),
and then retrieves the value from the future.

To minimize waiting,
the compiler pushes signal operations on each future
as far to the left into its producer conjunct as possible,
and it pushes wait operations
as far to the right into each of its consumer conjuncts as possible.
This means not only pushing them
into the body of the predicate called by the conjunct,
but also into the bodies of the predicates they call,
with the intention being that
each signal is put immediately after
the primitive goal that produces the value of the variable,
and each wait is put immediately before
the leftmost primitive goal that consumes the value of the variable.

\subsection{ThreadScope}

ThreadScope was originally built
to help programmers visualize the parallel execution of Haskell programs
compiled with the dominant implementation of Haskell,
the Glasgow Haskell Compiler or GHC.
The idea is that during the execution of a parallel Haskell program,
the Haskell runtime system writes time-stamped reports
about significant events to a log file.
The ThreadScope tool later reads this log file,
and shows the user graphically what each CPU was doing over time.
The diagrams it displays reveal to the programmer
the places where the program is getting the expected amount of parallelism,
as well as the places where it isn't.

We could adapt the ThreadScope system to Mercury for two main reasons.
The first is that the parallel implementations of Haskell and Mercury
in GHC and mmc (the Melbourne Mercury compiler)
are quite similar in several important respects,
even though they use different terms for
(slightly different implementations of) the same concepts.
For example, what Mercury calls an engine GHC calls a \emph{capability},
and what Mercury calls a context GHC calls a \emph{thread}.
(We will use Mercury terminology in the rest of the paper.)
The second reason is that
the ThreadScope log file format was designed to be extensible.
Each log file starts with a description
of each kind of event that may occur in it.
This description lists the arguments of the event and their total size,
which may be variable.
This description is precise enough that tools 
can skip events, and even of events, that they do not understand,
and process the rest.

Mercury is able to make use of a number of event types already supported by
ThreadScope and GHC,
in other cases we are able to add support for Mercury-specific event types to
ThreadScope.
Here we introduce the events in the first category,
The next section describes those in the second category.
Events of all types have a timestamp
that records the time of their occurrence, measured in nanoseconds.
This unit illustrates the level of precision ThreadScope aims for,
although of course there is no guarantee
that the system clock is capable of achieving it.

Most events in the log file are associated with an engine.
To avoid having to include an engine id with every event in the log file,
the log file format
groups sequences of events that are all from the same engine into a block,
and writes out the engine id just once, in a block header pseudo-event.
Since tools reading the log file can trivially remember
the engine id in the last block header they have read,
this makes the current engine id
(the id of the engine in whose block an event appears)
an implicit parameter of most event types.

The event types supported by the original version of ThreadScope 
that are relevant to our work are the following.

\emph{STARTUP:
Marks the beginning of the execution of the program,
and records the number of engines that the program will use.}

\emph{SHUTDOWN:
Marks the end of the execution of the program.}

\emph{CREATE\_THREAD:
Records the act of the current engine creating a context,
and gives the id of the context being created.}
(What Mercury calls a context Haskell calls a thread;
the name of the event uses Haskell terminology.)
Context ids are allocated sequentially,
but the parameter cannot be omitted,
since the runtime system can and will reuse the storage of a context
after the termination of the computation that has previously used it.

\emph{RUN\_THREAD:
Records the scheduling event
of the current engine switching to execute a context,
and gives the id of the context being switched to.}

\emph{STOP\_THREAD:
Records the scheduling event
of the current engine switching away from executing a context.}
Gives the id of the context being switched from,
as well as the reason for the switch.
The possible reasons include:
(a) the heap is full, and so the engine must invoke the garbage collector;
(b) the context has blocked, and
(c) the context has finished.

\emph{THREAD\_RUNNABLE:
Records that the current engine has made a blocked context runnable,
and gives the id of the newly-runnable context.}

\emph{RUN\_SPARK:
Records that the current engine is starting to run a spark
that it retrieved from its own local spark queue.}
Gives the id of the context that will execute the spark,
although this can be inferred by context.

\emph{STEAL\_SPARK:
Records that the current engine will run a spark
that it stole from the spark queue of another engine.}
Gives the id of the context that will execute the spark,
although again, this can be inferred by context.
Also gives the id of the engine that the spark was stolen from.

\emph{CREATE\_SPARK\_THREAD:
Records the creation of a new context or the reuse of an old context
in order to execute a spark.}
Gives the id of the context.
Does not say whether the context is new or reused.
In most cases, that information is simply not needed,
but if it is, it can be inferred from context.

\emph{GC\_START:
The current engine has initiated garbage collection;
control of the engine has been taken away from the mutator.}

\emph{GC\_END:
Garbage collection has finished on the current engine;
control of the engine has been returned to the mutator.}

\noindent
The longer a parallel program being profiled runs,
the more events it will generate.
Long running programs can generate enormous log files.
To keep the sizes of log files down as much as possible,
events include only the information they have to.
If some information about an event
can be inferred from information recorded for other events,
then the design principles of ThreadScope say
that information should be inferred at analysis time
rather than recorded at profiling runtime.
(The presence of context ids in RUN\_SPARK and STEAL\_SPARK events
violates this principle,
since they are guaranteed to be the same as the id of the context already
running on that engine or
the context id in the following CREATE_SPARK_THREAD event.
but their removal is prevented by backwards compatibility concerns.)
In most cases, the required inference algorithm is quite simple.
For example, answering the question of whether the context
mentioned by a CREATE\_SPARK\_THREAD event is new or reused
merely requires searching the part of the log up to that event
looking for mentions of the same context id.
And we have already seen how the engine id parameter
missing from most of the above events
can be deduced from block headers.

\section{New events}
\label{sec:newevents}

In order to support the profiling of parallel Mercury programs,
we had to add new arguments to two of the existing ThreadScope events.
Originally, the RUN\_SPARK and STEAL\_SPARK events
did \emph{not} specify the identity of the spark being run or stolen.
This is not a problem for the Haskell version of ThreadScope,
since it does not care about sparks' identities.
However, we do, since sparks correspond to conjuncts in parallel conjunctions,
and we want to give to the user not just general information
about the behavior of the program as a whole,
but also about the behavior of individual parallel conjunctions,
and of the conjuncts in them.
We have therefore added an id that uniquely identifies each spark
as an additional argument of the RUN\_SPARK and STEAL\_SPARK events.
Note that CREATE\_SPARK\_THREAD does not need a spark id,
since the only two ways it can get the spark it converts into a context
is by getting it from its own queue or from the queue of another engine.
A CREATE\_SPARK\_THREAD event will therefore always be preceded
by either a RUN\_SPARK event or a STEAL\_SPARK event,
and the id of the spark in that event
will be the id of the spark being converted.

We have extended the set of ThreadScope event types
to include several new types of events.
Most of these record information about constructs that do not exist in Haskell:
parallel conjunctions, conjuncts in those conjunctions, and futures.
Some provide information about the behavior of Mercury engines
that ThreadScope for Haskell does not need,
either because that information is of no interest,
or because the information is of interest
but it can be deduced from other events.
Even though the Haskell and Mercury runtime systems
generate many of the same events,
the stream of these common events they generate
do not necessarily obey the same invariants.

However, most of the work we have done towards adapting ThreadScope to Mercury
has been in the modification of the parallel Mercury runtime system
to generate all of the existing, modified and new events when called for.

In the rest of this section,
we report the event types we have added to ThreadScope.
In the next section, section \ref{sec:analysis},
we will show how the old and new events can be used together
to infer interesting and useful information
about the behavior of parallel Mercury programs.

\emph{START\_PAR\_CONJUNCTION: records the fact that
the current engine is about to start executing a parallel conjunction.}
It identifies the parallel conjunction in two different ways.
The static id identifies
the location of the conjunction in the source code of the program.
The dynamic id is the address of the barrier structure
that all the conjuncts in the conjunction will synchronize on when they finish.
Note that barrier structures may be garbage collected,
and their storage reused for other barriers used by later code,
so these dynamic ids are not unique across time,
but they \emph{do} uniquely identify a parallel conjunction
at any given moment in time.
See the end of this section for a discussion of why we chose this design.

\emph{END\_PAR\_CONJUNCTION:
records the end of a parallel conjunction}.
It gives the dynamic id of the finishing conjunction.
Its static id can be looked up in the matching START\_PAR\_CONJUNCTION event.

\emph{CREATE\_SPARK: records the creation of a spark
for a parallel conjunct by the current engine.}
Gives the id of the spark itself,
and the dynamic id of the parallel conjunction the conjunct is from.
To keep the log file small,
it does \emph{not} give the position of the conjunct in the conjunction.
However, since in every parallel conjunction
the spark for conjunct $n+1$
will be created by the context executing conjunct $n$,
and unless $n=1$, that context will itself have been created
from the spark for conjunct $n$.
(The first conjunct of a parallel conjunction is always executed directly,
without ever being represented by a spark.)
This means that the sparks for the non-first conjuncts
will always be created in order,
which makes it quite easy to determine which spark represents which conjunct.

The id of a spark is an integer consisting of two parts.
The first part is the number of the engine that created the spark,
and the second is an engine-specific sequence number,
so that in successive sparks created by the same engine,
the second part will be 1, 2, 3 etc.
This design allows the runtime system
to allocate globally unique spark ids without synchronization.

\emph{END\_PAR\_CONJUNCT:
records that the current engine
has finished the execution of a conjunct in a parallel conjunction.}
Gives the dynamic id of the parallel conjunction.
Note that there is no START\_PAR\_CONJUNCT event
to mark the start of execution of any conjunct.
For the first conjunct, its execution will start
as soon as the engine has finished recording the START\_PAR\_CONJUNCTION event.
The first thing the conjunct will do is create a spark
representing the second and later conjuncts
(which above we informally referred to as the spark for the second conjunct).
The CREATE\_SPARK event records the id of the spark,
then, either the RUN_SPARK or STEAL_SPARK event records which engine and context
executes the spark,
If the engine doesn't yet have a context a CREATE_SPARK_THREAD event is posted
that identifies the newly created context.
RUN\_THREAD and STOP\_THREAD events
also tell us when that context is executing, and on which engine.
The first thing the second conjunct does is spawn off a spark
representing the third and later conjuncts.
By following these links,
a single forward traversal of the log file can find out,
for each engine, which conjunct of which dynamic parallel conjunction
it is running at any given time
(if in fact it is running any;
an engine can be idle,
or it may be running code that is outside of any parallel conjunction).
When an engine records an END\_PAR\_CONJUNCT event,
it can only be for the conjunct it is currently executing.

\emph{FUTURE\_CREATE: records the creation of a future by the current engine.}
Gives the name of the variable the future is for,
as well as the dynamic id of the future.
It does not give the dynamic id of the parallel conjunction
whose conjuncts the future is intended to synchronize,
but that can be found out quite easily:
just scan forward for the next START\_PAR\_CONJUNCTION event.
The reason why this inference works is that
the code that creates futures
is only ever put into programs by the Mercury compiler,
and mmc never puts that code anywhere
except just before the start of a parallel conjunction.
If a parallel conjunction uses $n$ futures,
then every one of its START\_PAR\_CONJUNCTION events
will be preceded by $n$ FUTURE\_CREATE events,
one for each future.

The dynamic id of the future is its address in memory.
This has the same issues with reuse
as the dynamic ids of parallel conjunctions.

\emph{FUTURE\_SIGNAL: records that code running on the current engine
has signalled the availability
of the value of the variable protected by a given future.}
Gives the id of the future.
If another conjunct is already waiting for the value stored in this future,
this event will unblock its context.

\emph{FUTURE\_WAIT\_NO\_SUSPEND: records that the current engine
retrieved the value of the future without blocking.}
Gives the id of the future.
The future's value was already available
when the engine tried to wait for the value of the future,
so the context running on the current engine was not blocked.

\emph{FUTURE\_WAIT\_SUSPEND: records that the current engine
tried to retrieve the value of a future, but was blocked.}
Gives the id of the future.
Since the future's value was not already available,
the current context has been be suspended until it is.

We have added a group of four event types
that record the actions of engines
that cannot continue to work on what they were working before,
either because the conjunct they were executing finished,
or because it suspended.

\emph{TRY_GET_RUNNABLE_CONTEXT: This engine is checking the global
run queue for a context to execute.}

\emph{TRY_GET_LOCAL_SPARK: This engine is attempting to get a spark
from its own stack.}

\emph{TRY_STEAL_SPARK: This engine is attempting to steal a spark
from another engine.}

\emph{ENGINE_WILL_SLEEP: The engine is about to sleep.}
The next event from this engine will be from when it is next awake.

The idea is that an idle engine can look for work in three different places:
(a) the global runnable context queue,
(b) its own local spark queue, and
(c) another engine's spark queue.
An idle engine will try these in some order,
the actual order depending on the scheduling algorithm.
The engine will post the try event for a queue
before it actually looks for work in that queue.
If one of the tests is successful, the engine will post
a START_THREAD event, a RUN_SPARK event or a STEAL_SPARK event respectively.
If one of the tests fails, the engine will go on the next.
If it does not find work in any of the queues,
it will go to sleep after posing the ENGINE_WILL_SLEEP event.

This design uses two events to record
the successful search for work in a queue:
TRY_GET_RUNNABLE_CONTEXT and START_THREAD,
TRY_GET_LOCAL_SPARK and RUN_SPARK,
or TRY_STEAL_SPARK and STEAL_SPARK.
This may seem wasteful compared to using one event,
but this design enables us to measure several interesting things:
\vspace{-2mm}
\begin{itemize}
\item
how quickly a sleeping engine can wake up and try to find work
when work is made available by a CREATE_SPARK and or CONTEXT_RUNNABLE event;
\item
how often an engine is successful at finding work;
\item
when it is successful,
how long it takes an engine to find work and begin its execution;
\item
when it is unsuccessful,
how long it takes to try to find work from other sources,
fail, and then to go to sleep.
\end{itemize}
\vspace{-2mm}

\noindent
As we mentioned above,
using the address of the barrier structure
as the dynamic id of the parallel conjunction whose end the barrier represents
and using the address of the future structure
as the dynamic id of the future
are both design choices.
The obvious alternative would be to give them both sequence numbers
using either global or engine-specific counters.
However, this would require
adding a field to both structures to hold the id, which costs memory, and
filling in the field, which costs time.
Both these costs would be incurred by the program execution
whose performance we are trying to measure,
interfering with the measurement itself.
While such interference cannot be avoided
(writing events out to the log file also takes time),
we nevertheless want to minimize it if at all possible.
In this case, it is possible:
the addresses of the structures are trivially available,
and require no extra time or space during the profiled run.
The tradeoff is that we now need a pre-pass over the log file
that consistently renames apart
the different incarnations of the same id for a parallel conjunction or future.

\section{Deriving metrics from events}
\label{sec:analysis}

ThreadScope is primarily used to visualize the execution of parallel programs.
However, we can also use it, along with our new events,
to calculate and to present to the user
a number of different metrics about the execution of parallel Mercury programs.
Some of the metrics are of interest
only to application programmers,
or only to runtime system implementors,
or only to autoparallelization tool implementors,
but several are useful to two or even all three of those audiences.
Also, some of these metrics say something about the whole program;
some say something about individual parallel conjunctions,
and some say something about individual conjuncts inside those conjunctions.
We discuss our proposed metrics in that order;
we then discuss how we intend to present them to users.

\subsection{Whole program metrics}

\emph{CPUS\_OVER\_TIME:
The number of CPUs being used by the program at any given time.}
It is trivial to scan all the events in the trace,
keeping track of the number of engines currently being used,
each of which corresponds to a CPU.
The resulting curve tells programmers
which parts of their program's execution is already sufficiently parallelized,
which parts are parallelized but not yet enough to use all available CPUs,
and which parts still have no parallelism at all.
They can then focus their efforts on the latter.
The visualization of this curve has been implemented,
and is described by \cite{threadscope}.

\emph{GC\_STATS: The number of garbage collections,
and the average, minimum, maximum and variance
of the elapsed time taken by each collection.}
Mercury uses the Boehm-Demers-Weiser garbage collector~\cite{boehm}.
This collector supports parallel marking using GC-specific helper threads
rather than the OS threads that run Mercury engines.
Therefore, even when parallel marking is used,
we can only calculate the elapsed time used by garbage collection,
and not the \emph{cpu time} used by garbage collection.
The elapsed time of each garbage collection
is the interval between pairs of GC\_START and GC\_END events.

\emph{MUTATOR\_VS\_GC\_TIME:
The fraction of the program's runtime used by the garbage collector.}
We calculate this by summing the times
between pairs of GC\_START and GC\_END events,
and dividing the result by the program's total runtime.
(Both sides of the division refer to elapsed time, not CPU time.)
Due to Amdahl's law~\cite{amdahlslaw},
this fraction limits the best possible speedup we can get
for the program as a whole by parallelizing the mutator.
For example, if the program spends one third of its time doing GC
(which unfortunately actually happens for some parallel programs),
then no parallelization of the program can yield a speedup of more than three,
\emph{regardless} of the number of CPUs available.

\emph{NANOSECS\_PER\_CALL:
The average number of nanoseconds between successive procedure calls.}
The Mercury profiler for sequential programs~\cite{conway:2001:mercury-deep}
(which has nothing to do with ThreadScope)
measures time in call sequence counts (CSCs);
the call sequence counter is incremented at every procedure call.
It does this because there is no portable, or even semi-portable way
to access any realtime clocks that may exist on the machine,
and even the non-portable method on x86 machines (the RDTSC instruction)
is too expensive for it.
(The Mercury sequential profiler needs to look up the time at every call,
which in typical programs will happen every couple of dozen instructions or so.
The Mercury parallel profiler needs to look up the time only at each event,
which occur \emph{much} more rarely.)

The final value of the call sequence counter
at the end of a sequential execution of a program 
gives its length in CSCs; say $n$ CSCs.
When a parallelized version of the same program is executed on the same data
under ThreadScope,
we can compute the total amount of \emph{user time}
taken by the program on all CPUs; say $m$ nanoseconds.
The ratio $n/m$ gives the average number of nanoseconds
in the parallel execution of the program
per CSC in its sequential execution.
This is useful information,
because our automatic parallelization system
uses CSCs, as measured by the Mercury sequential profiler,
as its unit of measurement of the execution time 
of both program components and system overheads.
Using this scale, we can convert predictions about time made by the tool
from being expressed in CSCs to being expressed in nanoseconds,
which is an essential first step in comparing them to reality.

\subsection{Conjunction specific metrics}

\emph{PARCONJ\_TIME: The time taken by a given parallel conjunction.}
For each dynamic parallel conjunction id that occurs in the trace,
it is easy to compute the difference between
the times at which that parallel conjunction starts and ends,
and it is just as trivial to associate these time intervals
with the conjunction's static id.
From this, we can compute,
for each parallel conjunction in the program that was actually executed,
both the average time its execution took,
and the variance in that time.
This information can then be compared,
either by programmers or by automatic tools,
with the sequential execution time of that conjunction recorded by
Mercury's sequential profiler,
to see whether executing the conjunction was a good idea or not.
\emph{provided} that the two measurements are done in the same units.
At the moment, they are not, but as we discussed above,
CSCs can be converted to nanoseconds using the NANOSECS_PER_CALL metric.

\emph{PARCONJ\_RUNNABLE\_SELF:
The number of CPUs that can be used by a given parallel conjunction.}
For each dynamic parallel conjunction id that occurs in the trace,
we can divide the time between its start and end events into blocks,
with the blocks bounded by
the CREATE\_SPARK\_THREAD and STOP\_THREAD events for its conjuncts,
and the FUTURE\_WAIT\_SUSPEND and FUTURE\_SIGNAL events
of the futures used by those conjuncts.
We can then compute, for each of these blocks,
the number of runnable tasks that these conjuncts represent.
From this we can compute the history of the number
of runnable tasks made available by this dynamic conjunction.
We can derive the maximum and the time-weighted average of this number,
and we can summarize those
across all dynamic instances of a given static conjunction.

Consider a conjunction with $n$ conjuncts.
If it has a maximum number of runnable tasks
that is significantly less than $n$,
this is a sign that the parallelism the programmer aimed for
could not be achieved.
If instead the average but not the maximum number of runnable tasks
is significantly less than $n$,
this suggests that the parallelism the programmer aimed for was achieved,
but only briefly.
The number of runnable tasks can drop due to
either a wait operation that suspends or the completion of a conjunct.
Both dependencies among conjuncts
and differences in the execution times of the conjuncts
limit the amount of parallelism available in the conjunction.
The impact of the individual drops on the time-weighted average
shows which effects do the most limiting in any given parallel conjunction.

\emph{PARCONJ\_RUNNABLE\_SELF\_AND\_DESC:
The number of CPUs that can be used by a given parallel conjunction
and its descendants.}
This metric is almost the same as PARCONJ\_RUNNABLE\_SELF,
but it operates
not just on a given dynamic parallel conjunction,
but also on the parallel conjunctions
spawned by the call-trees of its conjuncts as well.
It takes their events into account when it divides time into blocks,
and it counts the number of runnable tasks they represent.

The two metrics,
PARCONJ\_RUNNABLE\_SELF and PARCONJ\_RUNNABLE\_SELF\_AND\_DESC,
can be used together to see whether
the amount of parallelism that can be exploited
by the two parallel conjunctions together is substantially greater than
the amount of parallelism that can be exploited
by just the outer parallel conjunction alone.
If it is, then executing the inner conjunction in parallel is a good idea;
if it is not, then it is a bad idea.

If the outer and inner conjunction in the dynamic execution
come from different places in the source code,
then acting on such conclusions is relatively straightforward.
If they represent different invocations of the same conjunction in the program,
which can happen if one of the conjuncts contains a recursive call,
then acting on such conclusions will typically require
the application of some form of runtime granularity control.

\emph{PARCONJ\_RUNNING\_SELF:
The number of CPUs that are actually used by a given parallel conjunction.}
This metric is computed similarly to PARCONJ\_RUNNABLE\_SELF,
but it doesn't count a runnable conjunct until gets to use a CPU,
and stops counting a conjunct when it stops using the CPU
(when it blocks on a future, and when it finishes).

Obviously, PARCONJ\_RUNNING\_SELF can never exceed PARCONJ\_RUNNABLE\_SELF
for any parallel conjunction at any given point of time.
However, one important difference between the two metrics
is that PARCONJ\_RUNNING\_SELF can never exceed
the number of CPUs on the system either.
If the maximum value of PARCONJ\_RUNNABLE\_SELF for a parallel conjunction
does not exceed the number of CPUS,
then its PARCONJ\_RUNNING\_SELF metric
can have the same value as its PARCONJ\_RUNNABLE\_SELF metric,
barring competition for CPUs by other parallel conjunctions (see later)
or by the garbage collector.

On the other hand, some conjunctions
do generate more runnable tasks than there are CPUs.
In such cases, we want the extra parallelism
that the system hardware cannot accommodate
in the period of peak demand for the CPU
to ``fill in'' later valleys,
periods of time when the conjunction demands
less than the available number of CPUs.
This will happen only to the extent that
the delayed execution of tasks that lost the competition for the CPU
does not lead to further delays in later conjuncts
through variable dependencies.

The best way to measure this effect
is to visually compare the PARCONJ\_RUNNING\_SELF curves for the conjunction
taken from two different systems,
e.g.\ one with four CPUs and one with eight.
However, given measurements taken from e.g.\ a four CPU system,
it should also be possible to predict with \emph{some} certainty
what the curve would look like on an eight CPU system,
by using the times and dependencies recorded in the four-CPU trace
to simulate how the scheduler would handle the conjunction
on an eight CPU machine.
Unfortunately, the simulation cannot be exact
unless it correctly accounts for \emph{everything},
including cache effects and the effects on the GC system.

The most obvious use of the PARCONJ\_RUNNING\_SELF curve of a conjunction
is to tell programmers whether and to what extent
that parallel conjunction can exploit the available CPUs.

\emph{PARCONJ\_RUNNING\_SELF\_AND\_DESC:
The number of CPUs that are actually used by a given parallel conjunction
and its descendants.}
This metric has the same relationship to PARCONJ\_RUNNING\_SELF
as PARCONJ\_RUNNABLE\_SELF\_AND\_DESC has to PARCONJ\_RUNNABLE\_SELF.
Its main use is similar to the main use of PARCONJ\_RUNNING\_SELF:
to tell programmers whether and to what extent
that parallel conjunction and its descendants can exploit the available CPUs.
This is important because a conjunction and its descendants
may have enough parallelism
even if the top-level conjunction by itself does not,
and in such cases the programmer can stop looking for more parallelism,
at least in that part of the program's execution timeline.

\emph{PARCONJ\_AVAIL\_CPUS:
The number of CPUs available to a given parallel conjunction.}
Scanning through the entire trace,
we can compute and record the number of CPUs being used
at any given time during the execution of the program.
For any given parallel conjunction,
we can also compute PARCONJ\_RUNNING\_SELF\_AND\_DESC,
the number of CPUs used by that conjunction and its descendants.
By taking the difference between the curves of those two numbers,
we can compute the curve of the number of CPUs
that execute some task \emph{outside} that conjunction,
Subtracting that difference curve
from the constant number of the available CPUs
gives the number of CPUs available for use by this conjunction.

This number's maximum, time-weighted average
and the shape of the curve of its value over time,
averaged over the different dynamic occurrences
of a given static parallel conjunction,
tell programmers the level of parallelism they should aim for.
Generating parallelism in a conjunction
that consistently exceeds the number of CPUs available for that conjunction
is more likely to lead to slowdowns from overheads
than to speedups from parallelism.

\emph{PARCONJ\_CONTINUE\_ON\_BEFORE:
the probability, for any given parallel conjunction,
that the code after the conjunction
will continue executing on the same CPU
as the code before the conjunction.}
If the parallel conjunction as a whole took only a small amount of time,
then CPUs other than the original CPU will still have relatively cold caches,
even if they ran some part of the parallel conjunction.
We want to keep using the warm cache of the original CPU.
The better the scheduling strategy is at ensuring this,
the more effectively the system as a whole will exploit the cache system,
and the better overall system performance will be.

\emph{PARCONJ\_CONTINUE\_ON\_LAST:
the probability, for any given parallel conjunction,
that the code after the conjunction
will continue executing on the same CPU
as the last conjunct to finish.}
If the parallel conjunction as a whole took a large amount of time,
then how warm the cache of a CPU will be
for the code after the conjunction
depends on how recently that CPU executed
either a conjunct of this conjunction or the code before the conjunction.
Obviously, all the conjuncts execute after the code before the conjunction,
so if the conjunction takes long enough for most of the data accessed
before the conjunction to 
be evicted from the cache,
then only the CPUs executing the conjuncts will have useful data in
their caches.
In the absence of specific information
about the code after the conjunct preferentially accessing data
that was also accessed (read or written) by specific conjuncts,
the best guess is that the CPU with the warmest cache
will be the one that last executed a conjunct of this conjunction.
The more often the scheduling strategy executes
the code after the conjunction on that CPU,
the more effectively the system as a whole will exploit the cache system,
and the better overall system performance will be.

\subsection{Conjunct specific metrics}

There are several simple times
whose maximums, minimums, averages and variances
can be computed for each static parallel conjunct.

\emph{CONJUNCT\_TIME\_AS\_SPARK:
the time between the conjunct's creation
(which will be as a spark) and the start of its execution
(when the spark will be converted into a context).}

\emph{CONJUNCT\_TIME\_AS\_CONTEXT:
the time between the start of the conjunct's execution and its end.}

\emph{CONJUNCT\_TIME\_BLOCKED:
the total amount of time
between the start of the conjunct's execution and its end
that the conjunct spends blocked waiting for a future.}

\emph{CONJUNCT\_TIME\_RUNNABLE:
the total amount of time
between the start of the conjunct's execution and its end
that the conjunct spends runnable but not running.}

\emph{CONJUNCT\_TIME\_RUNNING:
the total amount of time
between the start of the conjunct's execution and its end
that the conjunct spends actually running on a CPU.}

Since every context is always either running, runnable or blocked,
the last three numbers must sum up to the second
(in absolute terms and on average, not in e.g.\ maximum or variance).

Programmers may wish to look at conjuncts
that spend a large percentage of their time blocked
to see whether the dependencies that cause those blocks can be eliminated.

\emph{CONJUNCT\_TIME\_AFTER:
the time, for each parallel conjunct,
between the end of its execution and the end of the conjunction as a whole.}
It is easy to compute this for every dynamic conjunct,
and to summarize it as a minimum, maximum, average and variance
for any given static conjunct.
If the average times after
for different conjuncts in a given parallel conjunction
are relatively stable (have low variance)
but differ substantially compared to the runtime of the conjunction as a whole,
then the speedup from the parallel execution of the conjunction
will be significantly limited by Amdahl's law.
In such cases, the programmer may wish to take
two conjuncts that are now executed in parallel
and execute them in sequence.
Provided the combined conjunct is not the last-ending conjunct,
and provided that delaying the execution of one of the original conjuncts
does not unduly delay any other conjuncts that consume the data it generates,
the resulting small loss of parallelism
may be more than compensated for
by the reduction in parallelism overhead.
It is of course much harder and maybe impossible
to select the two conjuncts to execute in sequence
if the times after for at least some of the conjuncts
are \emph{not} stable (i.e. they have high variance).

\emph{FUTURE_SUSPEND_TIME_FIRST_WAITS:
for each shared variable in each parallel conjunction:
the minimum, average, maximum and variance of the time
of the first wait event on a future for this variable
by each consuming conjunct
minus the time of the corresponding signal event.}
If this value is consistently positive,
then contexts never or rarely suspend waiting for this future;
if this value is consistently negative,
then contexts often suspend on this future, and do so for a long time.
Programmers should look at shared variables
that fit into the second category
and try to eliminate the dependencies they represent.
If that is not possible, they may nevertheless try to reduce them
by computing the variable earlier
or pushing the point of first consumption later.

\emph{FUTURE_SUSPEND_TIME_ALL_WAITS:
for each future in each dynamic parallel conjunction:
the minimum, average, maximum and variance of the time
of all wait events minus the time of the corresponding signal event,
and the count of all such wait events.}
This metric helps programmers understand how hard
delaying the point of first consumption of a shared variable in a conjunct
is likely to be.
Delaying the first consumption is easiest
when the variable is consumed in only a few places,
and the first consumption occurs a long time before later consumptions.
If the variable is consumed in many few places,
many of these points of consumption are just slightly after
the original point of first consumption,
then significantly delaying the point of first consumption
requires eliminating the consumption of the shared variable
in \emph{many} places in the code,
or at least significantly delaying the execution
of those many pieces of code.

\emph{FUTURE_SUSPEND_PROBABILITY_FIRST_WAIT:
for the first wait event of each future:
the number of times the signal event occurs before the wait event versus
the number of times the wait event occurs before the signal event.}

\emph{FUTURE_SUSPEND_PROBABILITY_ALL_WAITS:
for each wait event of each future,
the number of times the signal event occurs before the wait event versus
the number of times the wait event occurs before the signal event.}
Like FUTURE_SUSPEND_TIME_ALL_WAITS and FUTURE_SUSPEND_TIME_FIRST_WAIT,
these two metrics can help programmers find out
how often contexts are suspended waiting for futures.

\emph{WAITED_FUTURE_SIGNAL_TO_CONJUNCT_END_TIME:
the average, maximum and variance of the time
between the signaling of a future on which another context is blocked
and the end of the parallel conjunct that signalled the future.}
This should be computed for every parallel conjunct
that signals at least one future.
If the average is below the time that it normally takes
for a context made runnable to begin executing on an engine,
and the variance is low, then it suggests that
it is better to execute the context made runnable by the signal
on the same engine immediately after the current conjunct finishes.

\emph{FUTURE_SIGNAL_TO_CONJUNCT_END_TIME:
the average, maximum and variance of the time
between the signaling of a future
and the end of the parallel conjunct that signaled that future.}
As above, this should be computed for
every parallel conjunct that signals at least one future.
This value can be used to determine
how often the optimization described above will not be useful.

\emph{OUT\_OF\_ORDER\_PROBABILITY:
given two parallel conjuncts $A$ and $B$,
with $A$ coming logically before $B$
either by being to the left of $B$ in some conjunction,
or by some ancestor of $A$ being to the left of an ancestor of $B$
in some conjunction,
how much time does the system spend with $B$ running
while $A$ is runnable but not running,
compared to the time it spends with either $A$ or $B$ running?}
Ideally, when $A$ and $B$ are both runnable
but the system has only one available CPU,
it should choose to run $A$.
Since it comes logically earlier,
the consumers that depend on its outputs
will also tend to come logically earlier.
Delaying the running of $A$ has a substantial risk of delaying them,
thus delaying the tasks depending on \emph{their} outputs, and so on.
Any efficient scheduling algorithm
must take this effect into consideration.

\emph{OUT_OF_ORDER_SCHEDULING_BLOCKS_ANOTHER_CONTEXT:
when tasks are executed out of order, as described above,
how much of the time is another context $C$
blocked on a future produced by $A$?}
This metric describes how often
out of order execution has a direct impact on another task.

\emph{OUT_OF_ORDER_SCHEDULING_BLOCKS_ANOTHER_CONTEXT_TC:
When tasks are executed out of order, as described above,
how much of the time do other contexts $D, E, F\ldots$ block
waiting on a future signaled by $C, D, E, \ldots$
which, eventually, depend on $A$?}
This metric considers the transitive closure of the dependency chain
measured by the previous metric.

\subsection{Presenting metrics to users}

The ThreadScope tool is a graphical program.
Its main display screen shows
a continuous timeline of the program's execution on the horizontal axis,
while along the vertical axis,
the display is divided into a discrete number of rows,
with one row per CPU (per engine in Mercury, per capability in Haskell).
The color of a display point for CPU $n$ at time $t$
shows what CPU $n$ was doing at time $t$:
whether it was idle, doing GC, or executing the program.
The time axis can be scaled to see an overview of the execution as a whole
or to focus on a specific period during the program run.
If the current scale allows for it,
the display also shows the individual events
that the on-screen picture is derived from.
Above the per-CPU rows ThreadScope shows a plot that is virtually identical
to the curve of the CPUS\_OVER\_TIME metric we defined above,
and now it can do so for Mercury programs as well as for Haskell programs.

The main ThreadScope GUI also has a mechanism that allows the user
to ask for simple scalar data about the profiled program run.
Originally, it was used only to display the overall time taken by the run.
It is trivial to modify this to let the user input
the number of call sequence counts (CSCs)
executed by a sequence version of the same program on the same data,
which would allow the tool to compute and display
the value of the NANOSECS\_PER\_CALL metric.
It is not much harder to modify it to compute and then output
the values of the GC\_STATS and MUTATOR\_VS\_GC\_TIME metrics.

We plan to modify the ThreadScope GUI so that
if the user hovers the pointer over the representation
of an event connected with a parallel conjunction as a whole
(such as a START\_PAR\_CONJUNCTION or END\_PAR\_CONJUNCTION event),
they get a menu allowing them to select
one of the conjunction-specific metrics listed above.
The system should then compute and print the selected metric.
Similarly if the user hovers the pointer over the representation
of an event connected with a specific parallel conjunct,
they should be able to ask for and get
the value of one of the conjunct-specific metrics.

We plan to add to ThreadScope a screen
that lists all the parallel conjunctions executed by the program.
(The information needed for that list is available in the strings
representing the static ids of the executed parallel conjunctions.)
By selecting items from this list,
the tool should be able to print summaries
from all the dynamic occurrences of the conjunction
for any of the conjunction-specific metrics.
It should also be able to take users
to all those occurrences in the main display.

We also plan to add to ThreadScope
a mechanism that allows the user to request
the generation of a plain text file in a machine-readable format
containing all the metrics that may be of interest
to our automatic parallelization tool.
We want our tool to be able to compare
its predictions of autoparallelized programs
with metrics derived from actual measurements of such programs,
so that we can help tune its performance to reduce the resulting discrepancies.

\section{Related work and conclusion}
\label{sec:conc}

Many parallel implementations of programming languages
come with visualization tools
since (as we argued in the introduction)
the people writing parallel programs
need such tools to understand the actual behavior of their programs,
and the people implementing those languages find them useful too.
Such tools have been created for all kinds of programming languages:
imperative (even Visual Studio has one),
functional
(\cite{edentraceviewer,loidl98:gransim,runciman93:profilingparfp} among others)
and logic (\cite{Foster96,vace}
and the systems cited by \cite{Gupta95parallelexecution}).

Many of these visualizers share common ideas and features.
For example, many systems (including ThreadScope)
use a horizontal bar with different parts colored differently
to display the behavior of one CPU,
while many other systems use trees
that display the structure of the computation.
(The latter approach is more common
among visualizers for sequential languages.)
Nevertheless, each visualizer is necessarily oriented
towards giving its users what they want,
and the users of different systems want different information,
since they are using different languages based on different concepts.
For example, functional languages have no concept of parallel conjunctions,
and their concept of data flow between computations
is quite different from Mercury's,
and users of concurrent logic programming languages
such as Concurrent Prolog, Parlog and Guarded Horn Clauses
have quite different concerns from users of Mercury
(where to disable parallelism, not where to enable it).

We know of only one visualizer
for a not-inherently-concurrent logic programming language
that does nevertheless support dependent AND-parallelism: VACE \cite{vace}.
However, the ACE system supports
OR-parallelism and independent AND-parallelism
as well as dependent AND-parallelism,
and its features don't seem designed
to help programmers exploit dependent AND-parallelism better.
As far as we know, noone has built a profiler
for a dependent AND-parallel logic programming language
with the capabilities that we propose.
We haven't yet built one either :-),
but we are working on it.
We expect to have at least a third of our proposed metrics implemented
by the time of the workshop,
and many more by the end of the year.

We would like to thank everyone
who has contributed to the development of ThreadScope,
in particular Simon Marlow and Duncan Coutts,
who have answered many of our questions,
and helped us add Mercury specific features to ThreadScope.

{\small
\bibliographystyle{eptcs}
\bibliography{tscope}
}

\end{document}